\documentclass[5p,sort&compress]{elsarticle}

\usepackage{hyperref}
\usepackage{graphicx}
\usepackage{xcolor}

\journal{Physica C}

\newcommand{\CBS}{Cu$_x$Bi$_2$Se$_3$~}
\newcommand{\SBS}{Sr$_x$Bi$_2$Se$_3$~}
\newcommand{\NBS}{Nb$_{0.25}$Bi$_2$Se$_3$~}

\begin{document}

\begin{frontmatter}

\title{An increase in $T_c$ under hydrostatic pressure in the superconducting doped topological insulator Nb$_{0.25}$Bi$_2$Se$_3$}

\author[ANL,UND]{M. P. Smylie\corref{cor1}}
\cortext[cor1]{Corresponding author: msmylie@anl.gov}
\author[ANL]{K. Willa}
\author[UIUC]{K. Ryan}
\author[ANL]{H. Claus}
\author[ANL]{W.-K. Kwok}
\author[MUST]{Y. Qiu}
\author[MUST]{Y. S. Hor}
\author[ANL]{U. Welp}

\address[ANL]{Materials Science Division, Argonne National Laboratory, Argonne, IL 60439}
\address[UND]{Department of Physics, University of Notre Dame, Notre Dame, IN 46556}
\address[UIUC]{Department of Physics, University of Illinois at Urbana-Champaign, Urbana, Illinois 61801}
\address[MUST]{Department of Physics, Missouri University of Science and Technology, Rolla, Missouri 65409}

\begin{abstract}
We report an unexpected positive hydrostatic pressure derivative of the superconducting transition temperature in the doped topological insulator \NBS via $dc$ SQUID magnetometry in pressures up to 0.6 GPa.
This result is contrary to reports on the homologues \CBS and \SBS where smooth suppression of $T_c$ is observed.
Our results are consistent with recent Ginzburg-Landau theory predictions of a pressure-induced enhancement of $T_c$ in the nematic multicomponent $E_u$ state proposed to explain observations of rotational symmetry breaking in doped Bi$_2$Se$_3$ superconductors.
\end{abstract}


\end{frontmatter}


\section{Introduction}
The pursuit of topological materials in condensed matter systems in recent years has initiated widespread research into the theoretical underpinnings as well as materials realizations of this new state of matter.
Following the prediction and discovery of topological insulators \cite{Kane-Mele-PRL-Z2-topological-order-and-QSHE,Hsieh-Hor-Cava-Nature-discovery-of-first-3D-TI,Hasan-Kane-RevModPhys-Review-of-TI,Qi-RevModPhys-Review-of-TSC-theory} in systems with strong spin-orbit coupling, much effort has been spent searching for topological superconductors.
In both, topological insulators and superconductors, a gap in the electronic band structure of the bulk coexists with symmetry-protected gapless states residing on the surface.
In the case of topological superconductors, these gapless surface states are predicted \cite{FuKane-PRL-TI-and-TSC} to host Majorana fermions, particles that act as their own antiparticle, and which are integral to fault-tolerant quantum computing \cite{Wilczek-QC,Kitaev-AnnPhys-QC,Merali-Nature-QC-disorder}.

Currently, two routes are being pursued towards the realization of topological superconductivity: via the proximity effect at the interface between certain semiconductors  displaying strong spin-orbit coupling and conventional superconductors \cite{Beenakker-AnnRevCMatt-Majoranas-in-TSCs}, or via chemical doping of topological insulators such as Bi$_2$Se$_3$, whose topologically non-trivial character has been experimentally confirmed by ARPES measurements \cite{Xia-NatPhys-BS-ARPES}.
Among the bulk materials, the most studied is the M$_x$Bi$_2$Se$_3$ (M=Cu, Sr, Nb) family \cite{Hor-Cava-PRL-CBS-discovery,Liu-JACS-SBS-discovery,Hor-NBS-arxiv} formed by intercalated doping of the parent Bi$_2$Se$_3$ compound, where the dopant ion is incorporated interstitially in the van der Waals gap between quintuple Bi$_2$Se$_3$ layers.
These superconducting materials, with $T_c$ around 3 K, have the trigonal $R\bar{3}m$ crystal structure innate to Bi$_2$Se$_3$.
As such, it was unexpected that all members of M$_x$Bi$_2$Se$_3$ exhibit a pronounced two-fold basal-plane anisotropy in various superconducting quantities such as the upper critical field, Knight shift and specific heat \cite{Matano-NatPhys-CBS,deVisser-SciRep,HHW-NatComm-SBS,Asaba-PRX-NBS}.
Recently, a nematic superconducting state was proposed \cite{Fu-PRB-CBS-theory,FuBerg-PRL-CBS-parity-theory,VenderbosFu-PRB-nematic-vs-chiral} to account for these experimental findings.
This state arises in a two-orbit model with odd-parity pairing and two-dimensional $E_u$ representation.
Odd-parity Cooper pairing is consistent with the recent observations of a nodal superconducting gap in \NBS \cite{Me-PRB-NBS-TDO,Me-arxiv-PRL-NBS-disorder}, with magnetization measurements \cite{Das-PRB-triplet-CBS}, with Andreev reflection spectroscopy \cite{Kurter-vanHartlingen-arXiv-NBS-ZBCP} and with the effects of pressure \cite{Bay-PRL-CBS-pressure,Nikitin-PRB-SBS-basal-anisotropy-pressure,Manikandan-arXiv-SBS-pressure-BCS} on the superconducting state.
Pressure-induced superconductivity \cite{Kirshenbaum-PRL-BS-pressure} is also observed in the parent compound Bi$_2$Se$_3$ and remarkably stable reemergent superconductivity \cite{Zhou-PRB-SBS-reemergent-SC-pressure} is observed in \SBS at high pressure, concurrent with a structural phase transition. 

Here, we report on the dependence of $T_c$ on hydrostatic pressure in single crystal Nb$_{0.25}$Bi$_2$Se$_3$.
Contrary to \CBS and Sr$_x$Bi$_2$Se$_3$, $T_c$ of \NBS continuously increases with pressure up to $\sim$0.6 GPa.  We discuss these results in terms of a nematic superconducting state of Nb$_{0.25}$Bi$_2$Se$_3$.

\section{Methods}
High quality samples of \NBS were grown by the method of Ref. \cite{Hor-NBS-arxiv}.
Crystals cleave naturally along the $ab$ plane of the $R\bar{3}m$ lattice and several $\sim$800x800x200 $\mu$m$^3$ crystals were cut from as-grown ingots; a photo of the crystal selected for pressure measurements is shown in the inset of Fig. 1.
The superconducting transitions of the crystals were screened using resistivity and magnetization measurements.
A home-built SQUID magnetometer with milli-Gauss background field \cite{Claus-Custom-SQUID} was employed for determining the superconducting transition temperature from field-cooled and zero-field-cooled magnetization curves.
The samples consistently showed $T_c \sim$ 3.4 K with complete superconducting shielding fraction.  

\begin{figure}
\includegraphics[width=1\columnwidth]{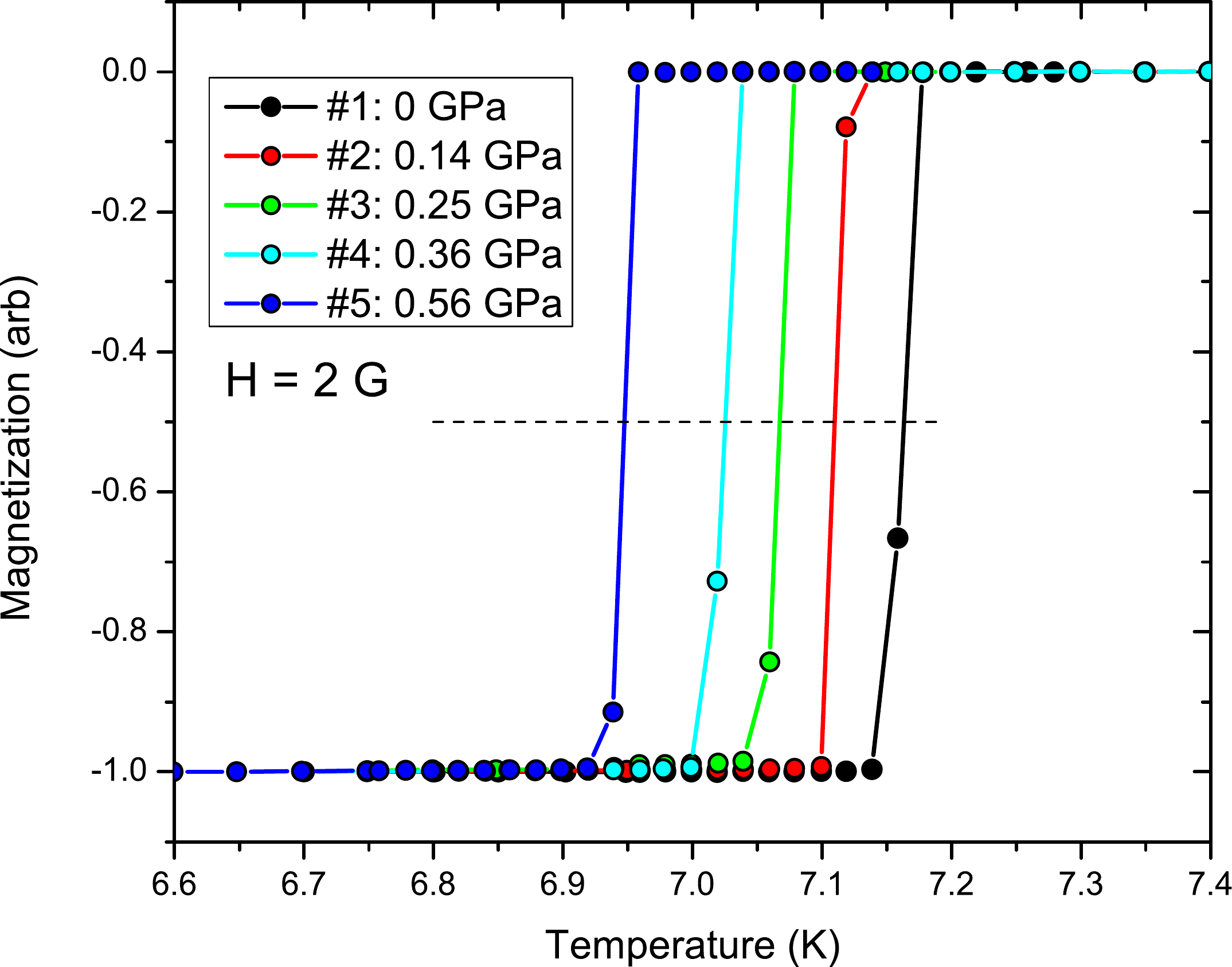}
\caption{
Normalized diamagnetic transitions in the polycrystalline Pb sample used as a manometer in the pressure cell under different applied hydrostatic pressures with H = 2 G.
Run \#1 refers to zero applied pressure.
From the values of the transition temperature, taken as the midpoint (dashed line) of the transition, we deduce the actual pressures of runs \#2 through \#5.
}
\label{fig1}
\end{figure}

Pressure dependent measurements were performed in a Quantum Design MPMS$\textsuperscript\textregistered$.
For these measurements, the remnant field in the superconducting magnet was carefully measured with a Pd reference sample and sufficient counter field was applied to reduce it to under 0.1 G in order to avoid spurious effects stemming from the strong field dependence of $T_c$ (see Fig. 3).
Pressure measurements were performed in a CuBe pressure cell, part of the Almax easyLab Mcell 10$\textsuperscript\textregistered$ attachment for the MPMS.
The pressure is applied hydrostatically in the CuBe Mcell by two small zirconia pistons, which compress a 3mm diameter Teflon capsule containing the sample and a proprietary nonmagnetic pressure transmitting medium.
Additionally, a small piece of high quality Pb was included in the capsule to act as a manometer.
In the Mcell 10 attachment, the pressure dependent superconducting transition curve of a small Sn sample is typically used to obtrain the pressure in the cell.
However, since the curve \cite{Sn-Tc-P-dependence} for Sn nearly overlaps with the transition of our \NBS crystals, we replaced Sn with a high-quality polycrystalline Pb sample.
Fig. 1 shows the normalized diamagnetic transition of Pb with increasing pressure from zero (black, \#1) to maximum (blue, \#5).
The transition decreases with pressure while the transition width remains narrow.
We estimate the pressure of the cell using a $T_c$ suppression rate \cite{Pb-Tc-P-dependence} of -0.386 K/GPa deduced from the Pb sample.
We note that the pressure dependence is reversible on increasing and decreasing pressure, with the recovery of the original $T_c$ following reduction to zero pressure.

For transport measurements, Au electrical contacts were evaporated onto freshly cleaved samples in a conventional 4-point arrangement, and Au-leads were attached with silver epoxy. 

\begin{figure}[t]
\includegraphics[width=1\columnwidth]{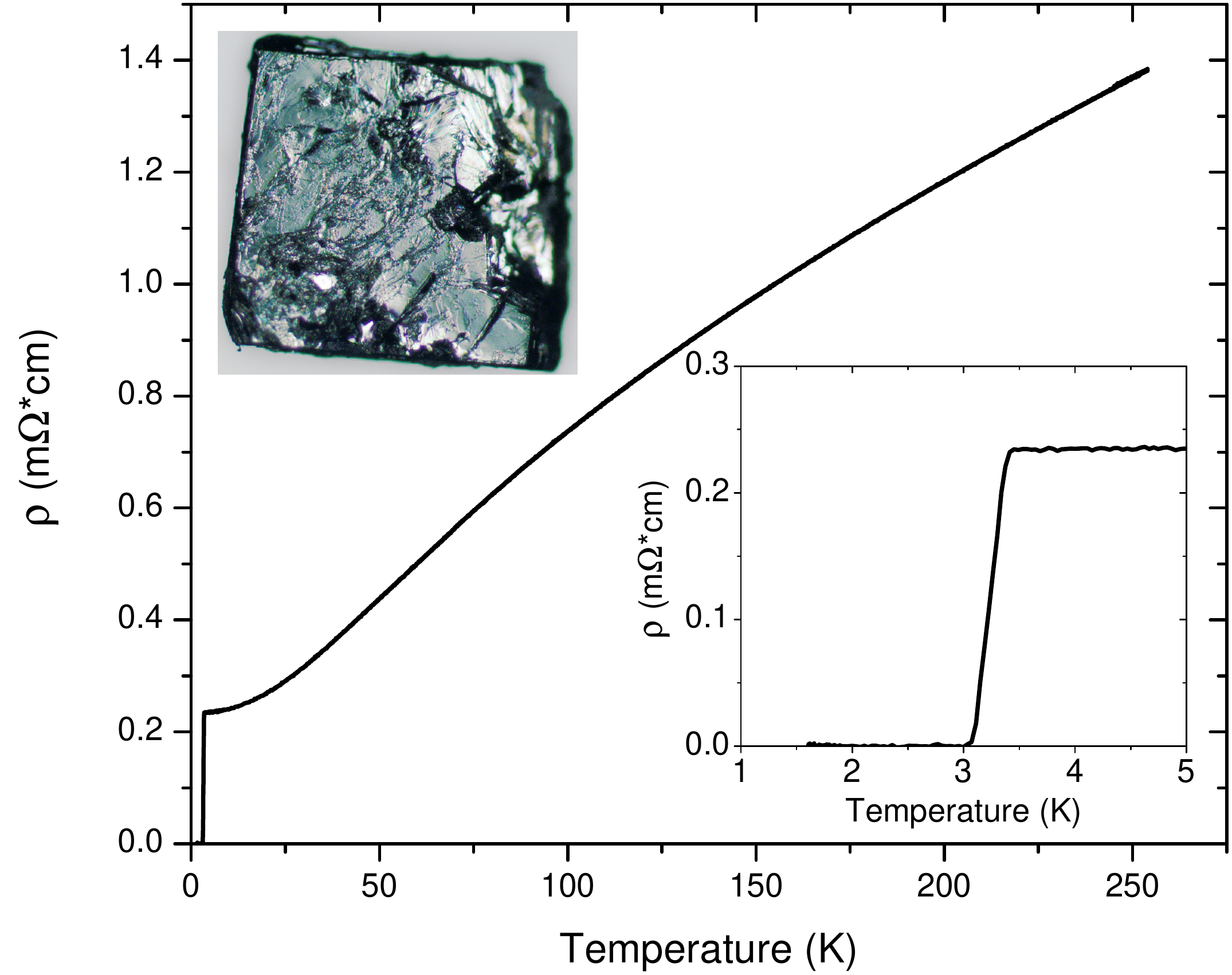}
\caption{
Resistivity of a single crystal of Nb$_{0.25}$Bi$_2$Se$_3$; results on all crystals are similar.
There is a sharp superconducting transition at $\sim$3.25 K.
The lower inset shows the superconducting transition on an expanded scale revealing a transition width $\Delta T_c < $ 0.5 K.
The upper inset shows a photograph of the crystal selected for pressure cell measurements.
}
\label{fig2}
\end{figure}

\section{Results}
Fig. 2 shows the temperature dependence of the resistivity of a 470x220x40 $\mu$m$^3$ crystal.
The material is metallic with a residual resistivity ratio R$_{250K}$/R$_{4K} \sim$ 6 and a fairly low residual resistivity, $\rho_0 \sim$ 0.25 m$\Omega *$cm.
The observation of quantum oscillations in similar crystals \cite{Lawson-PRB-dHvA-NBS} is taken as evidence of the high purity of this material.
Using a single-band Drude model, the electron mean free path $l$ is given as $l = m^* v_F / (n e^2 \rho)$.
Here, $m^*$, $v_F$, $e$, $\rho$ and $n$ are the effective mass, Fermi velocity, electron charge, resistivity, and carrier concentration, respectively.
Using $m^*$ = 0.15 and $v_F$ = 7x10$^5$ m/sec from dHvA measurements (average values from \cite{Lawson-PRB-dHvA-NBS}), $\rho_0$ = 0.25x10$^{-4}~\Omega$*cm from Fig. 1, and $n$ = 1.5x10$^{20}$ cm$^{-3}$ from Hall measurements \cite{Hor-NBS-arxiv, Lawson-PRB-dHvA-NBS}, we obtain $l \sim$ 50 nm.
Alternatively, the mean free path can also be estimated from $l$ = $v_F \tau.$
Using the average value for the scattering time of $\tau$ = 4.8x10$^{-14}$ s from \cite{Lawson-PRB-dHvA-NBS}, we find $l \sim$ 35 nm.
Due to the simplifications underlying the Drude model and to the uncertainty in the actual electronic structure of Nb$_{0.25}$Bi$_2$Se$_3$, these estimates should be considered as an order-of-magnitude estimate.
Nevertheless, experimental evidence suggests that \NBS is a clean superconductor such that $l > \xi$ with the coherence length $\xi \sim$ 19 nm as determined from the c-axis upper critical field \cite{Me-PRB-NBS-TDO}.

As $H_{c1}$ is only $\sim$75 G in this material \cite{Me-PRB-NBS-TDO}, careful measurement of the $T_c$ suppression in small applied fields was performed.
Results are shown in Fig. 3 for several fields with H // $c$.
For the pressure dependent measurements we chose a field of 2 G.

\begin{figure}
\includegraphics[width=1\columnwidth]{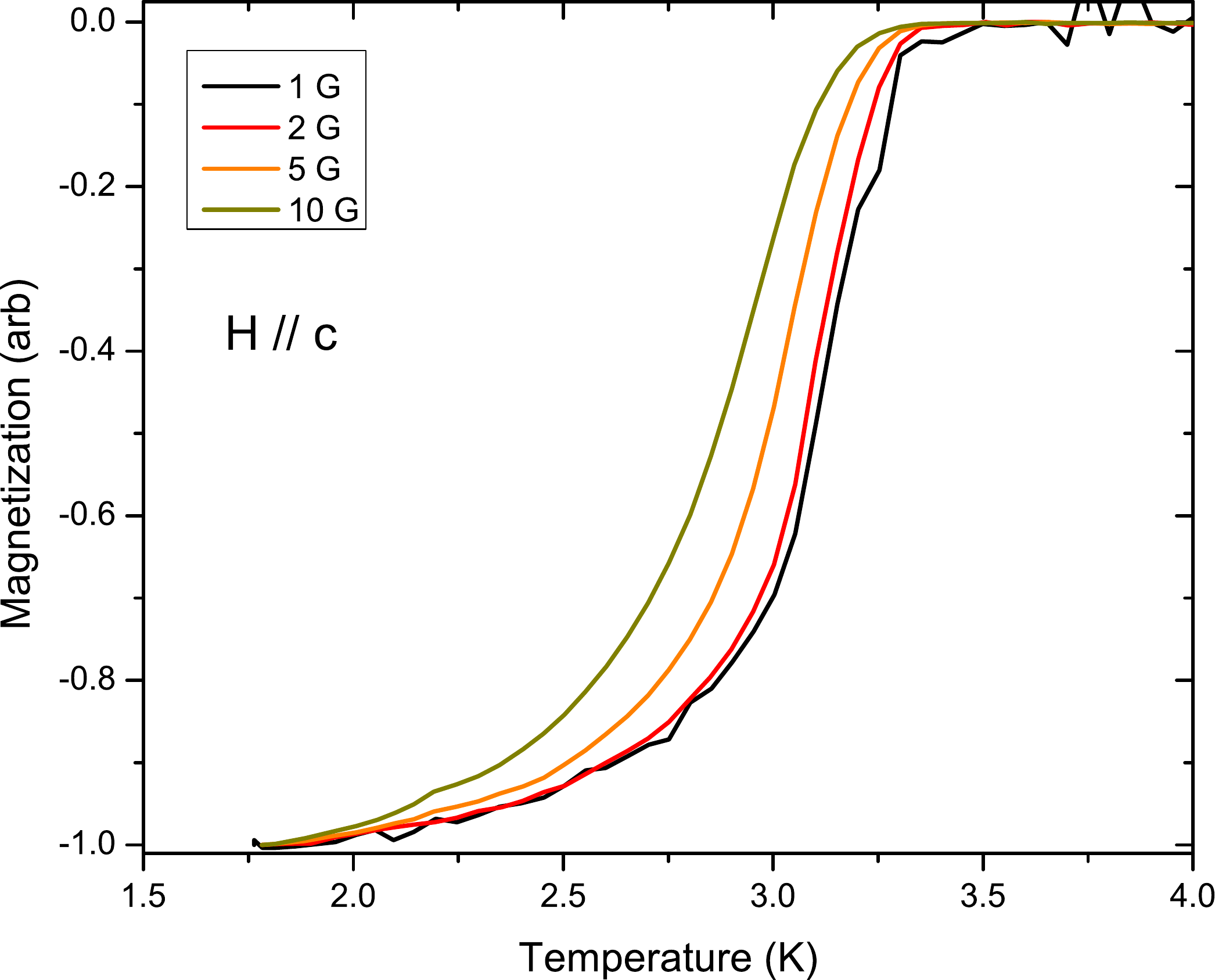}
\caption{
Normalized diamagnetic transitions in the single crystal of \NBS selected for pressure measurements in various applied fields with H // $c$. $T_c$ is rapidly suppressed with field. 
}
\label{fig3}
\end{figure}

Figure 4 shows the normalized diamagnetic transition of \NBS with increasing pressure from 0 to 0.56 GPa in applied fields of 2 G with H // $c$.
The superconducting transition shifts clearly to higher temperatures as pressure is increased, with an approximate rate of 0.47 K/GPa.
At the same time, the transition remains sharp and parallel.  

\begin{figure}
\includegraphics[width=1\columnwidth]{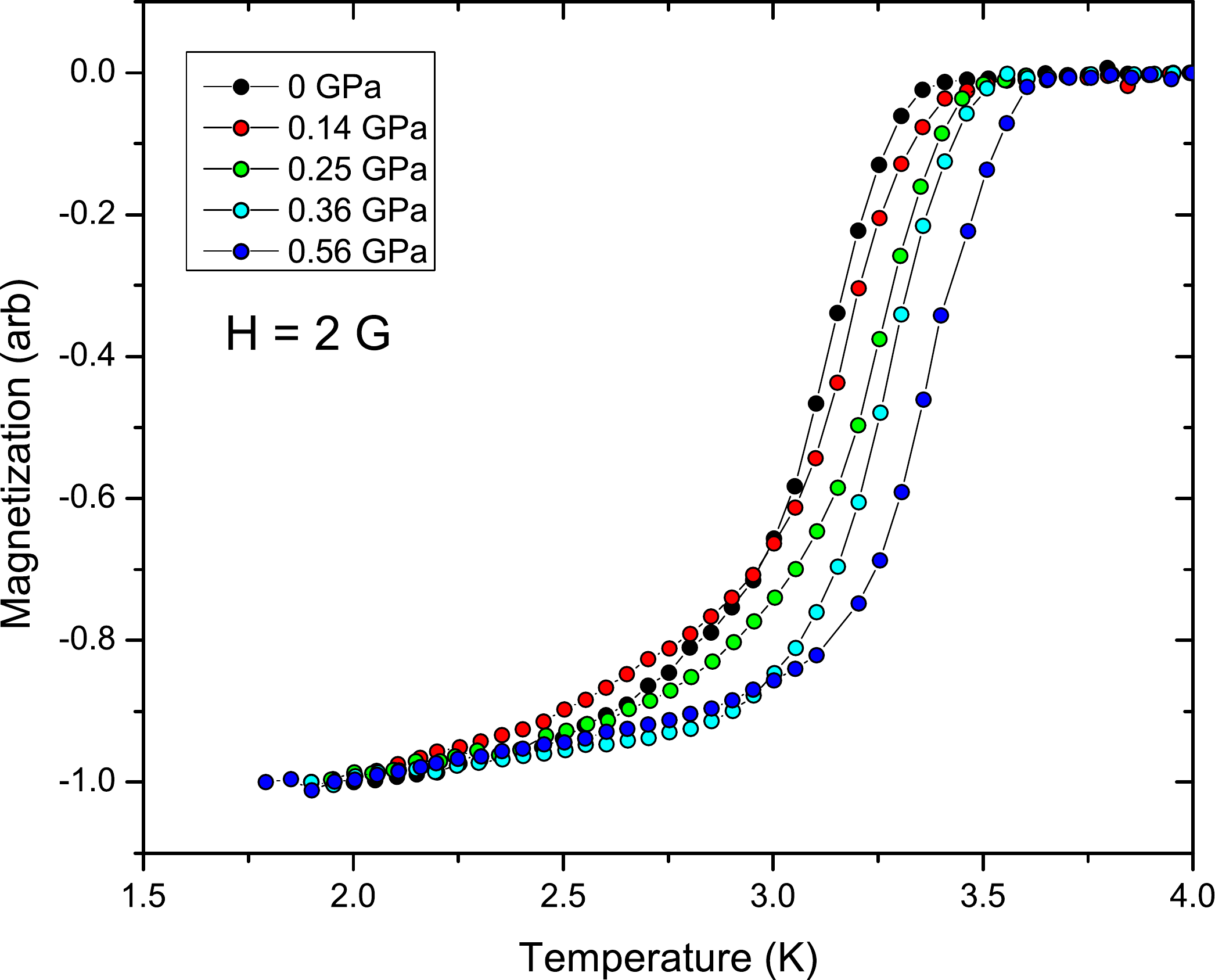}
\caption{
Diamagnetic transitions in a single crystal of \NBS with increasing pressure in H = 2 G with H // $c$.
There is noticeable increase in $T_c$ for all pressures.
}
\label{fig4}
\end{figure}

\section{Discussion}
The main result of our study is shown in Fig. 5, where we plot the percentage increase in $T_c$ (taken at the midpoint of the transition) as function of pressure, and compare it with values for related compounds obtained from the literature \cite{Bay-PRL-CBS-pressure,Nikitin-PRB-SBS-basal-anisotropy-pressure,Manikandan-arXiv-SBS-pressure-BCS}.
Our measurements demonstrate that \NBS is the only one with an initial positive d$T_c$/d$P$, while all others in this material class show negative d$T_c$/d$P$, with \SBS showing the fastest suppression at high pressure.
However, at low pressure, $T_c$ of \SBS is essentially pressure-independent (see Fig. 5) possibly displaying a (unresolved) very small enhancement in $T_c$ at the lowest pressures.
In a simple model for a low carrier density superconductor, the transition temperature will be dependent on the carrier concentration $n$ and the effective mass $m^*$ as $T_c \sim \Theta_D e^{1/N(0)V_0}$ where the density of states $N(0) \sim m^*n^{1/3}$ and $V_0$ is the effective interaction parameter \cite{Parks}.
The observation of an increase in $T_c$ suggests an increase in either or both the effective mass and the carrier concentration with pressure.
In the parent compound Bi$_2$Se$_3$, the carrier concentration increases with pressure \cite{Kirshenbaum-PRL-BS-pressure} leading to pressure-induced superconductivity.
In Cu$_x$B$i_2$Se$_3$, a decrease of the carrier concentration and thus $N(0)$ with pressure has been observed \cite{Bay-PRL-CBS-pressure}, and band structure analysis of \SBS suggests a decrease in the density of states with increasing pressure \cite{Manikandan-arXiv-SBS-pressure-BCS}.
An analysis based on Ginzburg-Landau theory \cite{Fu-PRB-CBS-theory} of the proposed unconventional $E_u$ pairing symmetry for the M$_x$Bi$_2$Se$_3$ family of superconductors suggests that there is a linear coupling between in-plane uniaxial strain and the superconducting order parameter, such that a small strain should enhance $T_c$.
The hydrostatic pressure derivative is given as $\frac{dT_c}{dP} = \Sigma_i \frac{\partial T_c}{\partial \sigma_i}$, where $\sigma_i$ is the stress in direction $i$.
Thus, within the model of a nematic superconductor, a net positive pressure derivative will arise as long as the $c$-axis uniaxial pressure derivative is not too large and negative.
Hence, this model will imply that the data presented in Fig. 5 represent large, negative $c$-axis uniaxial pressure derivatives for \CBS and Sr$_x$Bi$_2$Se$_3$.
We also note that quantum oscillation and ARPES measurements \cite{Lawson-PRB-dHvA-NBS, Lahoud-CBS-SdH} on \CBS and \SBS indicate a Fermi surface composed of a single elliptical sheet centered at the $\Gamma$-point whereas quantum oscillation measurements on \NBS give indication of additional small electron pockets which might be highly susceptible to pressure.

\begin{figure}
\includegraphics[width=1\columnwidth]{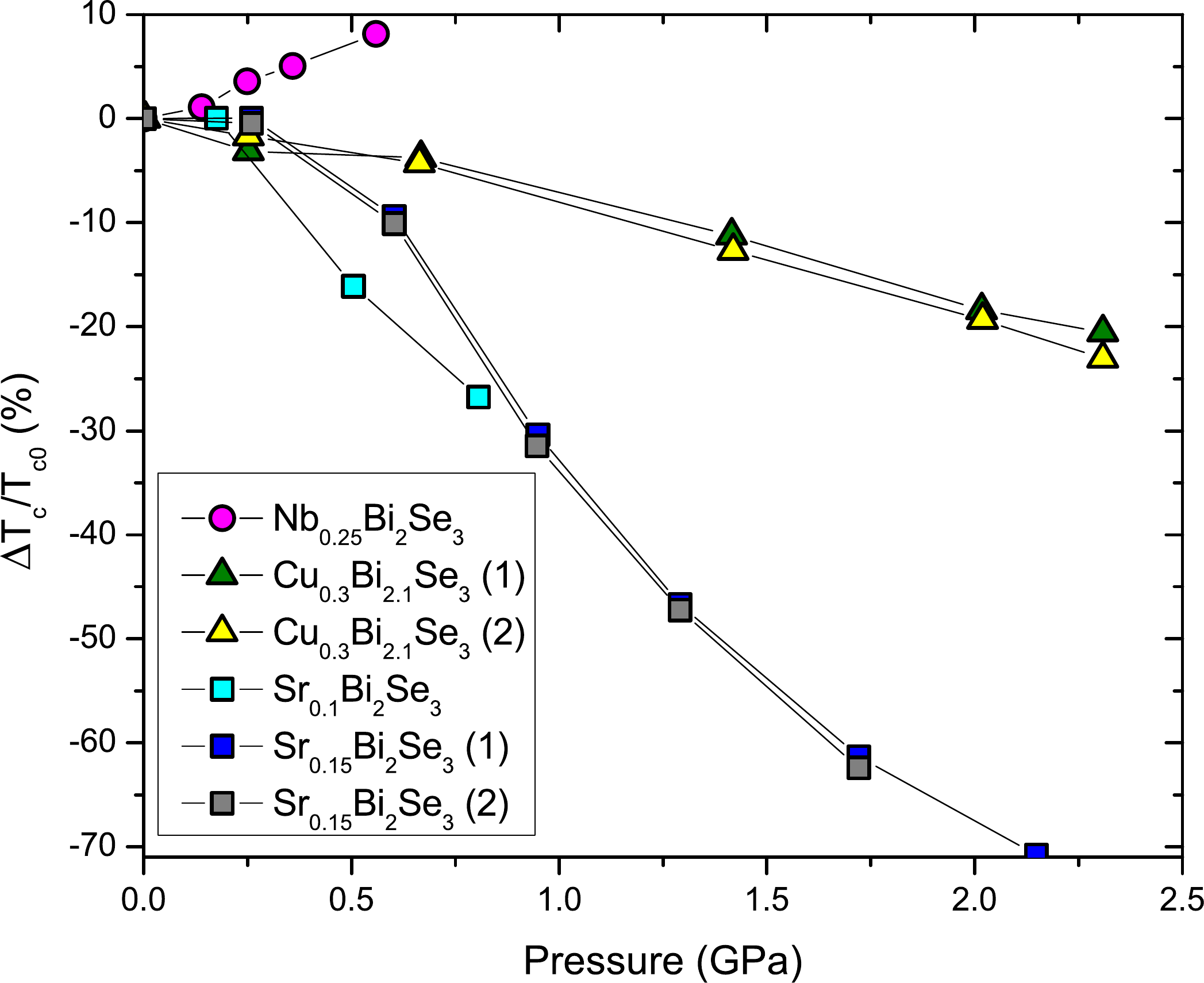}
\caption{
Comparison of fractional change in superconducting transition temperature vs applied hydrostatic pressure in multiple members of the superconducting M$_x$Bi$_2$Se$_3$ family, with data for M=(Cu,Sr) obtained from Refs. \cite{Bay-PRL-CBS-pressure,Nikitin-PRB-SBS-basal-anisotropy-pressure,Manikandan-arXiv-SBS-pressure-BCS}.
Unlike for M=(Cu,Sr), negative d$T_c$/d$P$ is not observed in M=Nb, and no trend towards saturation is approached within the measured range of applied pressure.
}
\label{fig5}
\end{figure}

\section{Summary}
The study of pressure-dependent effects on the superconducting state is a unique tool to probe for unconventional superconductivity.
Here, we presented the first finding of an unexpected positive pressure dependence of $T_c$ in Nb$_{0.25}$Bi$_2$Se$_3$, amid recent reports of smooth $T_c$ suppression with pressure in related materials \cite{Bay-PRL-CBS-pressure,Nikitin-PRB-SBS-basal-anisotropy-pressure,Manikandan-arXiv-SBS-pressure-BCS,Zhou-PRB-SBS-reemergent-SC-pressure}, indicating potential differences in the electronic structure of these materials.

\section{Acknowledgments}
Magnetization measurements were supported by the U.S. Department of Energy, Office of Science, Basic Energy Sciences, Materials Sciences and Engineering Division.
MPS thanks ND Energy for supporting his research and professional development through the ND Energy Postdoctoral Fellowship Program.
KW acknowledges support through an Early Postdoc Mobility Fellowship of the Swiss National Science Foundation.
YSH acknowledges support from the National Science Foundation grant number DMR-1255607.

\section*{References}

\bibliography{NBS_M_T_under_pressure_v0.5-arXiv}

\end{document}